\documentclass{iau}
\usepackage{graphicx}

\title[Ap stars in close binary systems] %% give here short title %%
{Candidate Ap stars in close binary systems}

\author[Folsom et al. ]   %% give here short author list %%
{C.P. Folsom$^{1,2}$, G.A. Wade$^{3}$, K. Likuski$^{3}$, O. Kochukhov$^{4}$, 
E. Alecian$^{5}$, D. Shulyak$^{6}$, N.M. Johnson$^{3}$}

\affiliation{
$^{1}$Institut de Recherche en Astrophysique et Plan\'etologie, Toulouse, France\\
email: {\tt colin.folsom@irap.omp.eu} \\
$^{2}$Armagh Observatory, Armagh, Northern Ireland\\
$^{3}$Department of Phyics, Royal Military College of Canada, Kingston, Canada\\
$^{4}$Department of Astronomy and Space Physics, Uppsala University, Uppsala, Sweeden\\
$^{5}$Observatoire de Paris, Meudon, France\\
$^{6}$Institute of Astrophysics, Georg-August-University, G\"ottingen, Germany\\
}

\pubyear{2013}
\volume{302}  %% insert here IAU Symposium No.
\pagerange{1--2}
\jname{Magnetic fields throughout stellar evolution}
\editors{P. Petit ... eds.}
\begin{document}

\maketitle

\begin{abstract}
Short period binary systems containing magnetic Ap stars are anomalously rare.  This apparent anomaly may provide insight into the origin of the magnetic fields in theses stars.  As an early investigation of this, we observed three close binary systems that have been proposed to host Ap stars.  Two of these systems (HD 22128 and HD 56495) we find contain Am stars, but not Ap stars.  However, for one system (HD 98088) we find the primary is indeed an Ap star, while the secondary is an Am star.  Additionally, the Ap star is tidally locked to the secondary, and the predominately dipolar magnetic field of the Ap star is roughly aligned with the secondary.  Further investigations of HD 98088 are planned by the BinaMIcS collaboration.  

%\keywords{Keyword1, keyword2, keyword3, etc.}
\end{abstract}

\firstsection % if your document starts with a section,
\section{Introduction}

Magnetic Ap stars in short period binary systems are very rare.  Whereas the incidence of other chemically peculiar A stars in close binary systems is at least as large as in single stars, the incidence of Ap stars in close binaries is much lower.  This observation may provide insight into the origin of magnetism in A-type stars, and is one of the avenues being pursued by the new Binarity and Magnetic Interactions in various classes of Stars (BinaMIcS) collaboration.  As an initial step in this project, we studied three close binary systems which have been suggested to contain Ap stars (HD 22128, HD 56495, \& HD 98088), in order to asses the presence of magnetic fields and study the atmospheric chemistry of the components.  High resolution spectropolarimetric observations of these stars were obtained with the MuSiCoS instrument at the Observatoire du Pic du Midi.  %in France.

\section{HD 98088}

HD 98088 is a SB2 (P = 5.905 days, e = 0.184; Carrier et al. 2002), which was identified as chemically peculiar by \cite{Abt1953}, and magnetic by \cite{Babcock1958}.  \cite{Carrier2002} studied the system's orbital parameters, but there are no modern magnetic or chemical abundance studies.  We present new results from \cite{Folsom2013a}.

We applied spectral disentangling to the set of Stokes $I$ observations of HD 98088.  The disentangled spectra were used for abundance analyses of the two components, by fitting them with synthetic spectra computed with the ZEEMAN code.  In the primary, we find strong overabundances of Fe-peak elements and rare earths, and roughly solar abundances for lighter elements, indicating an Ap star.  In the secondary, we find overabundances of Fe-peak elements and underabundances of Ca and Sc, indicating an Am star.  

In order to assess stellar magnetic fields, we performed Least Squares Deconvolution (LSD) on our observations, producing `mean' line profiles.  In Stokes $V$, we find clear Zeeman signatures in the primary's lines, and no signal in the secondary's lines.  For the primary, longitudinal magnetic fields were measured from the LSD profiles.  We measured the rotation period of the primary from the magnetic variability, and the result agrees well with the orbital period, implying that the system is tidally locked.  We find the magnetic field of the primary is predominately dipolar, with a polar strength of $3850 \pm 450$ G. 

Comparing the magnetic and orbital geometries, we find that one magnetic pole of the primary always points roughly towards the secondary.  The dipole axis appears to be $15 \pm 5^{\circ}$ out of the orbital plane, and thus the alignment may not be perfect, but it is very suggestive.

\section{HD 22128}

HD 22128 is a SB2 (P = 5.086 days, e $\sim$0; Carrier et al. 2002).  The primary was proposed as an Ap star by \cite{Olsen1979} and by \cite{Abt1979}.  Here we present new results from \cite{Folsom2013b}.

An abundance analysis was performed by fitting synthetic SB2 spectra computed with ZEEMAN to the observations.  In both stars, we find overabundances of Fe-peak elements and rare earths, and underabundances of Ca and Sc, indicating both components are Am stars.  We extracted LSD profiles from our observations, and find no detection of a magnetic signature in any Stokes V profile. Measuring longitudinal magnetic fields from these profiles we find no detection, with uncertainties of $\pm 50$ G in the primary, and $\pm 90$ G in the secondary.  

We conclude that HD 22128 is a close binary containing two very similar Am stars, but neither star is a magnetic Ap star.

\section{HD 56495}

HD 56495 is a SB2 (P = 27.38 days, e = 0.165; Carrier et al. 2002).  The primary was proposed as an Ap star based on a marginal magnetic detection by \cite{Babcock1958}.  We present new results from \cite{Folsom2013b}.

We performed an abundance analysis, fitting the SB2 spectrum with ZEEMAN.  For the primary, we find clear overabundances of Fe-peak elements and underabundances of  Ca and Sc, indicating an Am star.  For the secondary we find abundances consistent with solar.  We computed LSD profiles for our observations, and there is no detection of a magnetic signature in the Stokes V profiles.  Measuring longitudinal magnetic fields from the unblended profiles we detect no magnetic field, with uncertainties of $\pm 80$ G in the primary, and $\pm 100$ G in the secondary.  

We conclude that the primary is an Am star and the secondary a normal F star.  The results for HD~22128 and HD~56495 suggest that Ap stars in close binaries may be even more rare than previously thought.

\end{document}